# Speculating on the Role of Media Architecture in Post-Disaster Rebuilding and Recovery: Insights from Architects and Interaction Designers

Berk Göksenin Tan*

Arçelik Research Center for Creative Industries (KUAR), Koç University, Istanbul/Turkey, btan20@ku.edu.tr

Oğuzhan Özcan

Arçelik Research Center for Creative Industries (KUAR), Koç University, Istanbul/Turkey, oozcan@ku.edu.tr

In post-disaster contexts, design is not only about rebuilding structures but also about reimagining how architecture can become a communicative medium that supports recovery, resilience, and collective memory. While recent studies have expanded the understanding of media architecture from aesthetic urban screens to participatory civic infrastructures, there remains limited empirical research on its potential role in post-disaster contexts. In particular, opportunities exist to explore how architecture and interaction design might speculate on media architecture's role in rebuilding and recovery efforts for post-disaster permanent housing, especially when conceptualizing disasters as active agents that reshape design processes. Following the Kahramanmaraş earthquake on February 6, 2023, we conducted two focus groups with architects and interaction designers in the case of Antakya, Turkey, building on affected residents' expectations for post-earthquake permanent housing. Our analysis revealed three critical dimensions of how future media architecture may support post-disaster housing: (1) as a facilitator of individuals' social connection to their community, (2) as an enabler of multispecies participation and collective efforts, and (3) as a mediator of heritage preservation and revival. With novel perspectives, we contribute a three-dimension lens for media architecture in permanent homes; a co-speculative, card-based process bridging resident insights and expert design; and ten situated speculative design ideas with implications for design of post-disaster permanent homes.

## 1 INTRODUCTION

Permanent homes in post-disaster contexts are not merely shelters; they are critical anchors that help re-establish social life, cultural continuity, and everyday resilience after disruption [12, 26]. Rebuilding is therefore not only about constructing safer physical structures, but also about recovering the relations, practices, and meanings that sustain community identity. In places like Antakya, where the 6 February 2023 Kahramanmaraş earthquakes devastated a city that is dense with layered heritage, permanent housing must be understood as both material and symbolic infrastructure for recovery. It is in this context that new digital and spatial technologies, capable of transforming buildings into communicative and participatory actors, raise important questions about how recovery can be mediated, extended, and sustained.

---

* Place the footnote text for the author (if applicable) here.

Over the past two decades, the field of media architecture has emerged as a critical area of inquiry into how digital technologies and architectural form intersect to shape civic, cultural, and social experiences [23, 34]. Early work often emphasized large-scale media façades and aesthetic urban screens, producing iconic landmarks in global cities [29]. More recent studies, however, have expanded the understanding of media architecture toward participatory civic infrastructures that foster dialogue, interaction, and community engagement in public space [6]. This shift foregrounds the role of architecture not merely as a static structure but as a medium for communication and exchange.

For instance, after the 2011 Christchurch earthquake, Wesener [33] showed that temporary physical spaces with media architecture elements can be resilience-building infrastructures. More to that, events like *LUXCITY* created temporary light-based installations, performances, and participatory projects that transformed the ruined city center into spaces of joy and collective gathering, highlighting temporary media architecture installations' opportunity for experimental, collaborative, creative place-making in post-disaster contexts [13]. Yet despite these examples in public space and temporary settings, we know little about how media architecture might contribute to the specific conditions of post-disaster permanent housing, where recovery must be sustained over time and balanced with participation and heritage continuity.

This research addresses this gap by asking what are the potential roles of media architecture to support the recovery processes when considered within post-disaster permanent housing of heritage-rich contexts such as Antakya. Centering around the case of the Kahramanmaraş earthquake on February 6, 2023, which caused unprecedented destruction across southern Turkey and northern Syria [19], the city of Antakya became a critical site for questioning how permanent housing could be reimagined. To explore this question, we conducted two focus group sessions with architects and interaction designers to deliver speculative ideas, building on earlier research that captured residents' expectations for their future homes after the earthquakes. Using card-based prompts derived from residents' insights, the sessions generated 26 initial ideas that we synthesized into ten refined concepts for embedding media architecture into everyday domestic and communal spaces, producing situated visions of rebuilding and recovery for the post-disaster permanent homes in the city.

The findings from the focus group sessions show how media architecture can be understood across three interrelated dimensions: facilitating social connection within communities, enabling multispecies participation and collective efforts, and mediating heritage preservation and revival. By articulating these empirical insights and conceptual framings, and by demonstrating a co-speculative method that bridges community and expert perspectives, this study contributes the field of media architecture for more situated and context-sensitive infrastructures of resilience that are capable of supporting recovery while attending to the social, ecological, and heritage complexities of rebuilding in Antakya.

## 2 RELATED WORK

Media architecture has shifted from spectacular urban screens toward situated, participatory infrastructures that scaffold civic life and everyday practices. Early debates on the aesthetics of participation frame media architecture as staging new modes of engagement and emancipation [9], while critiques of ocularcentrism argue for multisensory and embodied interaction [7]. Hybrid-resolution, playful facades and digital placemaking demonstrate how place-based, programmable media can enhance belonging and communal memory [14, 20]. Inside dwellings and semi-public interiors, "interior media architecture" links social media practices and everyday data to spatial decision-making [17], and neighbourhood-scale resilience work connects concepts, strategies, and examples that operate locally [24]. Across these strands, content strategy matters: media systems should "inform, influence, and intrigue," aligning communicative purpose with situated context rather than spectacle [2].

Post-disaster scholarship adds requirements around safety, participation, and cultural fit. "Build Back Better" (BBB) reconceptualizes reconstruction to reduce vulnerability and improve governance [22], while People-Centered Housing Recovery (PCHR) emphasizes residents' agency and culturally grounded practices in permanent housing [21]. Temporary and transitional urbanism in Christchurch, such as LUXCITY's light-based installations and community programming, illustrate how low-threshold, time-bounded interventions reactivate public life and collective identity [13, 33]. For heritage-rich settings, European work foregrounds authenticity, co-creation, and participatory engagement to sustain continuity through change [8]. In parallel, minimally invasive, projection-based techniques support layered, accessible storytelling without compromising conservation priorities [32].

A complementary body of digital and immersive heritage research extends this beyond the site. Co-designed immersive storytelling has surfaced multi-vocal narratives and resilience in earthquake-affected Italian towns [10], and VR reconstructions have helped rebuild sense of place, enable intergenerational memory transmission, and support sensitive mourning practices [11]. Together, these trajectories outline a continuum: on-site, low-maintenance media layers embedded in thresholds, courtyards, facades, and shared rooms, complemented by higher-fidelity immersive experiences that deepen narrative work when appropriate.

Two gaps remain for permanent homes. First, convergence; prior work rarely examines how domestic-scale media architecture can simultaneously facilitate micro-civic coordination and care, enable collective stewardship across human and more-than-human actors, and mediate heritage preservation and revival over long horizons. Second, governance and care; beyond initial deployment, little is said about content moderation, community curation, repair, maintenance, and budget-wise responsibility, which are issues that are decisive for everyday, post-disaster living. In heritage-dense contexts, a sensitive approach is also necessary to avoid re-triggering while still enabling remembrance and learning.

Positioned against these gaps, our study focuses on post-disaster permanent homes as socio-technical media infrastructure. We shift attention from city-icon projects to intimate scales and embedded placements, broadening "users" to include residents, professionals, and more-than-human actors (e.g., buildings, ecologies), and emphasizing multisensory, robust modalities compatible with heritage sensitivities. In doing so, we extend BBB/PCHR principles into the everyday: aligning safety and participation with content governance and long-term stewardship, and translating heritage continuity into non-invasive, participatory media practices suited to Antakya's layered fabric.

## 3 RESEARCH CONTEXT AND METHOD

### 3.1 The Research Context and Background

On February 6, 2023, a magnitude 7.8 earthquake struck southern and central Turkey as well as northern Syria, with its epicenter located north of Kahramanmaraş. The disaster caused widespread destruction across the region and resulted in more than 52,000 casualties. Later the same day, a second earthquake of magnitude 7.6 intensified the damage, followed by a third shock of magnitude 6.5 on February 20 near Antakya, approximately 100 kilometers from the initial epicenter [19]. Together, these seismic events severely affected more than ten cities. Hatay Province, and particularly its capital Antakya, suffered the most devastation, where the shaking reached a maximum Mercalli intensity of XII. Government reports indicate that in Hatay alone, more than 847,000 residential structures and over 1,100 cultural heritage buildings were damaged or destroyed [1].

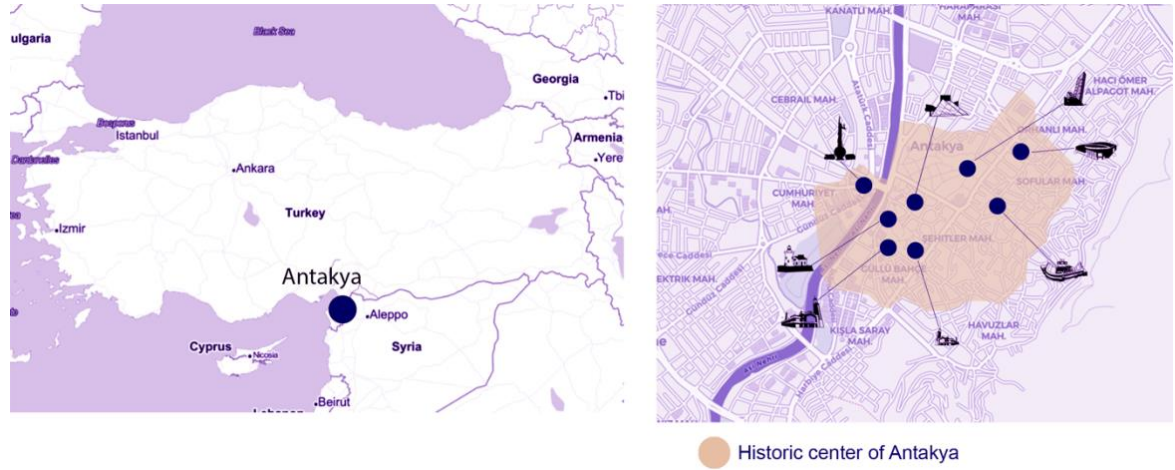

Figure A. Map of Antakya, showing the historic center

Antakya's dense layering of historical, cultural, and social complexity makes it one of the most challenging contexts for post-disaster permanent housing (see figure A). The city's history dates to 300 BC, and over time, it hosted many civilizations from Byzantium to the Ottomans, which continuously shaped and reshaped its fabric [15]. This long process of transformation has led to the coexistence of structures from different periods and the preservation of diverse traditional practices that contribute to its tangible and intangible heritage values [27, 31]. Yet, despite these challenges, large-scale rebuilding efforts were rapidly initiated in the aftermath of the catastrophe. In this accelerated recovery process, the need for more user-centered approaches to permanent housing became increasingly apparent. To address this gap, we conducted a qualitative study with affected residents, focusing on their expectations for post-earthquake domestic environments. Semi-structured interviews reported elsewhere [30], revealed four critical design themes: the importance of togetherness and community, the role of individual agency and efforts, the need to support economic recovery and regeneration, and the preservation of heritage values and memory. Each of these themes was supported by insights from the residents' accounts, ranging from the necessity of communal and collaborative spaces that enhance security, to opportunities for empowering individuals in technology use and reconstruction efforts, to strategies for enabling self-sufficiency and domestic production, and finally to approaches for re-experiencing and preserving cultural heritage in post-disaster life. These insights provided the conceptual foundation for the next stage of our study.

### 3.2  The Study Design and Analysis

Based on the themes identified with residents, we developed a set of ideation cards to facilitate structured discussion in expert workshops. Prior research has demonstrated the value of card-based methods as a medium for ideation and knowledge transfer, and our design followed these precedents [18, 25]. Each card included the name of the relevant theme, a color code for visual categorization, and a short insight distilled from the resident interviews. On the reverse side, an exemplary quotation from a resident was included to ground the insight in lived experience. This ensured that expert participants could not only engage with abstract design considerations but also encounter the voices of those directly affected. The appendix A illustrates an example of the cards used in the focus group studies.

We organized two expert focus group sessions, one with architects (n = 5) and one with interaction designers (n = 5) (as shown in Table 1). Participants were recruited through professional networks and were selected on the basis of their experience in practice or research, including prior engagement with post-disaster housing, participatory design, or interactive technologies. The sessions followed a replicated procedure. Each began with a short introduction to the research context and aims, followed by a warm-up discussion in which participants reflected on their own understanding and prior experiences of disaster and recovery contexts. The main activity was a card-based ideation and discussion exercise, during which the experts collectively speculated on possible futures for permanent housing in Antakya, particularly focusing on the potential role of media architecture as shown in figure B. Each session concluded with a collective synthesis phase in which emerging ideas were clustered and critically discussed.

| Focus Groups | Participants | Recruitment Expertise |
| --- | --- | --- |
| Session 1 | | |
| | P1 | She is a resident of Antakya who is affected by the earthquake. She practices as an architect in Antakya for more than 5 years. |
| | P2 | She practices architecture for more than 5 years, works in historical buildings preservation and restoration. |
| | P3 | He practices architecture for more than 5 years and designs historic buildings with adaptive reuse. |
| | P4 | She practices architecture for more than 5 years, mainly focuses on architectural design competitions. |
| | P5 | He practices architecture for more than 5 years and is an expert for restoration and renovation projects. |
| Session 2 | | |
| | P6 | She works in the field of human-building interaction, focusing on future of housing and urban design. |
| | P7 | She works in the field of more-than-human-centered design, focusing on plants and agricultural activities. |
| | P8 | He works in the field of human-computer interaction, focusing on embodied interaction with technologies. |
| | P9 | She works in the field of human-computer interaction, focusing on VR/AR spatial design and experiences. |
| | P10 | She works in the field of human-computer interaction, focusing on future domestic food practices. |

Table 1. Focus group participants

All sessions were audio-recorded with informed consent. Alongside session notes and the visual materials produced during the workshops, the recordings were transcribed and coded. Analysis followed a reflexive thematic analysis approach, which allowed us to iteratively move between the resident-derived themes and the expert-generated speculations [3] This analytic strategy made it possible to identify new dimensions and speculative directions at the intersection of architecture, interaction design, and post-disaster rebuilding, while remaining grounded in the perspectives and expectations of affected residents.

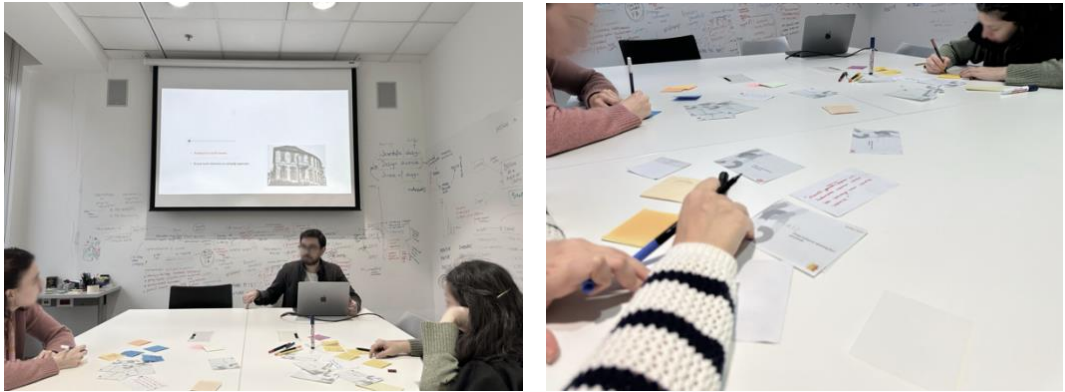

Figure B. Focus group sessions.

Following recent methodological practices in design research [4], we used a large language model (LLM, ChatGPT version 5) as an assistive tool to refine naming conventions and reduce duplication within the set of speculative ideas. No raw transcripts or sensitive data were shared with external services; only de-identified cluster summaries and candidate labels were processed. All outputs from the LLM were manually reviewed, refined, or rejected by the authors, and the provenance of AI-assisted contributions was fully documented. While the LLM facilitated organizational efficiency, all interpretive and conceptual decisions remained entirely human-driven to preserve reflexive analytical integrity to avoid interpretive bias and loss of nuance.

## 4 RESULTS

The two focus-group sessions produced 26 initial ideas, which we synthesized into ten refined ideas across three dimensions highlighting the potential role of media architecture in post-earthquake permanent homes in Antakya: (1) facilitating individuals' social connection to their community, (2) enabling multispecies participation and collective efforts, and (3) mediating heritage preservation and revival, as visually demonstrated in figure C. The refined ideas are presented as a synthesis of the overall discussions and speculative ideas within the focus group sessions, underlining the domain-specific contributions of architects and interaction designers.

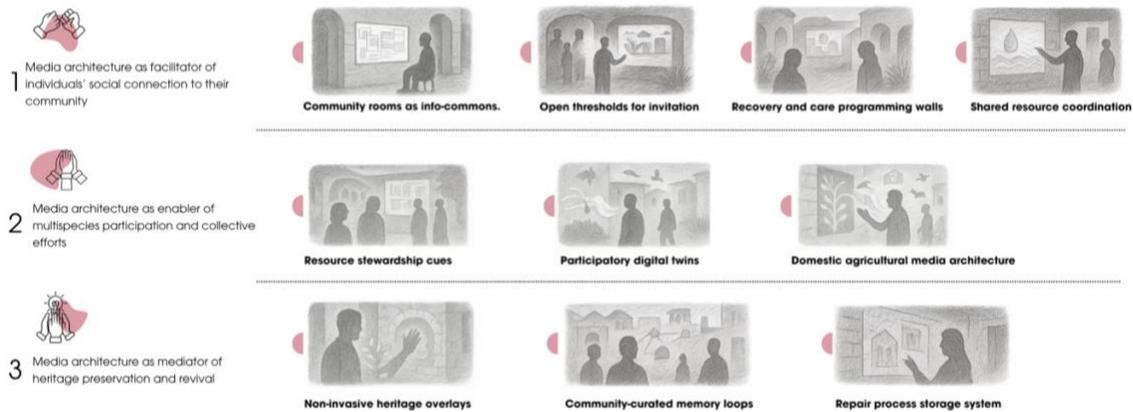

Figure C. The resulting dimensions and speculative design ideas, sketched by the authors with the support images generated with LLM.

### 4.1 Media architecture as facilitator of individuals' social connection to their community

During the focus group sessions, participants framed media architecture as a component of social infrastructure that supports everyday coordination and care while blending with Antakya's heritage fabric. Architects emphasized spatial logics such as ground-floor rooms, courtyards, and shared spaces, whereas interaction designers prioritized legibility, safety, and inclusion of residents. Together, they argued for conditions that sustain sociability, resisting gated or exclusionary post-earthquake architectural configurations and valued inclusive, safe information environments. For this theme we present four refined ideas:

The first speculative idea is **'Community rooms as info-commons.'** Participants ideated for semi-public interiors (courtyards, shared ground floors) act as low-power information commons where media-architecture elements communicate local news, events, aid distribution, and mutual-help sign-ups. Around the discussions of this idea, architects pointed out for spatial projections; and interaction designers underlined inclusive communication. The second idea is **'Open thresholds for invitation.'** Within the scope of this speculation, architectural thresholds (façades, foyers, arcades) are projected to be augmented with ambient media to invite neighbours into everyday encounters without reproducing gated arrangements; this was primarily articulated by architects advocating open public interfaces. From an architect's perspective, P1 highlighted the value of these spaces with media architecture elements that would potentially facilitate spending time together as a community. Another speculative idea is **'Recovery and care programming walls.'** This idea incorporated programmable façades and courtyard walls that host community-curated content to support individuals' recovery within collective life (e.g., support-group schedules, children's activities, shared rituals), combining provision for "rehab/care" spaces with collaborative content practices. While discussing this speculation, P6 remarked, "These wall could enable creating a circle for people who already know each other and extend it over time with inclusion of others, to get to find, connect or to know each other again... The feeling of security can also strengthen over time."

The last speculative idea of this theme is **'Shared resource coordination.'** Building-level media elements (e.g., simple dashboards) are speculated to render bookings and sharing for day-to-day operational needs (tools, laundry, water points, generator slots), strengthening reciprocity; interaction designers linked this directly to togetherness.

Taken together, these proposals position media architecture as an everyday, low-intensity social infrastructure for post-earthquake housing in Antakya. Instead of foregrounding display or spectacle, the four ideas prioritize calm communication embedded in shared rooms and thresholds, community-curated programming that supports recovery, and simple coordination tools that organize daily resources. The emphasis on spatial logics (courtyards, ground floors, shared areas) coupled with legibility, safety, and inclusion positions media architecture as a means to sustain sociability while preserving the relation with the city's heritage fabric. Therefore, this dimension frames media architecture as a layer for proximity, care, and reciprocity that helps residents rebuild collective life in the post-earthquake recovery process.

### 4.2 Media architecture as enabler of multispecies participation and collective efforts

Participants in both focus group sessions described collective rebuilding efforts as shared agency and participation distributed across humans, buildings, and environments. Discussions centered on affordability and safety as non-negotiable factors in the design of post-earthquake permanent homes in Antakya, positioning media architecture elements as more stable and economically implementable, rather than technically ambitious and hard-to-sustain solutions. Most spots of the discussion covered media architecture's potential for enabling equality and welfare of the community. Three refined speculative ideas in this focus are as follows:

The first idea of this theme is **'Resource stewardship cues.'** The speculation includes ambient feedback for resource use, such as water use, grey-water cycles, and rooftop-solar generation that supports a more self-sustaining built environment, encouraging collective conservation and awareness; particularly, interaction designers tied this to building-level agency. The other speculative idea of this theme is **'Participatory digital twins.'** This includes the use of digital twins surfaced through media-architecture elements in public and private spaces that support input for individual needs and collective decision-making in the post-earthquake homes of Antakya. Interaction designers also emphasized accommodating more-than-human stakeholders (e.g., stray animals) within these representations for the considerations of multispecies participation in the recovery process. The last speculative idea discussed in the focus group sessions is **'Domestic agricultural media architecture.'** The speculation includes vertical gardens and micro-green walls that act as communicative surfaces, guiding at-home soilless farming through light prompts and icons and encouraging product sharing. Architects focused on indoor planning constraints; interaction designers explored extending sharing and local know-how to neighbourhood and city scales. In relation to this idea, P4 noted, "Residential soilless farming could be an important implementation in this specific idea. Because it will be more feasible to integrate into the homes and easily become a part of the neighbourhoods' commercial activities."

Overall, in this dimension, speculations position media architecture as a pragmatic stewardship layer for post-earthquake life, where agency is shared across humans, buildings, and environments in the rebuilding process. Rather than proposing technically ambitious showpieces, participants emphasized more accessible and inclusive proposals that are easy to maintain for the residents. Across the three refined ideas, environmental phenomena become legible (through ambient cues on water and energy), actionable (through digital twins that convene collective decision-making and acknowledge more-than-human stakeholders), and livable (through domestic cultivation practices that can scale from the dwelling to the neighbourhood).

**4.3 Media architecture as mediator of heritage preservation and revival**

Given Antakya's historic background and sensitivity, focus group participants intersected on **respectful** strategies that prioritize preventing loss and enabling residents to protect the remaining tangible and intangible heritage values in the rebuilding process. Media architecture here is specifically framed not to deliver aesthetic reminder or representation of damage but to create situated, everyday encounters with heritage values aligned with conservation priorities and community expectations. We described three refined ideas below:

The first speculative idea in this theme is **'Non-invasive heritage overlays.'** It proposes using projection mapping and AR markers to re-surface lost façades, craft details and historic street alignments on surviving walls, specific architectural landmarks and streetscapes. Architects also noted its potential for preserving intangible heritage values (e.g., performing religious rituals in sacred architecture spaces), while interaction designers pointed out for integrating multisensory room-scale interaction (sound, olfactory and tactile cues) to evoke memory. For instance, while speculating non-invasive heritage overlays with multi-sensory interactions, P10 quoted as "It could be multi-sensory things like sound, smell, and texture. Sometimes, evoking a memory with a scent in the room-scale is very effective."

Idea 9 is **'Community-curated memory loops'** which includes public media walls that layer tangible and intangible heritage through memory loops and invite engagement by residents and visitors. Interaction designers specifically proposed these interactive walls to receive inputs and respond materially as people leave traces over time to enhance the interaction capability of this media architecture system. Idea 10 is **'Repair process storage system.'** A system that is city-scale, everyone connecting from their post-earthquake permanent homes with media-architecture elements to document and transmit tangible and intangible heritage, including craft-informed know-how central to rebuilding. This speculative idea also recognizes the agency of heritage values in shaping permanent homes. The discussions for the use of this system ranged from a digitally stored cookbook that has traditional recipes of Antakya to the craft knowledge for materials regarding traditional Antakya homes.

Altogether, these proposals cast media architecture within post-earthquake permanent homes as a restorative element for heritage values in the process of rebuilding Antakya. In most cases, residents re-encounter tangible and intangible traces in everyday setting, curate personal memories and transmit knowledge regarding heritage. With these speculations, media architecture mediates between past and future, care and re-construction: it enables learning, ritual, and transparency of the inherited values within the community, and recognizes heritage as an active agent shaping the design of permanent homes in Antakya.

## 5 DISCUSSION

### 5.1 The scale, placement, users and interaction modalities of media architecture in post-disaster contexts

The results from our focus group discussions about Antakya highlight that the *scale* and *placement* of media architecture are decisive in determining its efficacy in post-disaster permanent housing. Participants emphasized that interventions need to move beyond monumental or purely iconic installations toward neighbourhood and home scale applications embedded in facades, courtyards, and shared infrastructures, as well as distributed in public and private spatial contexts. This aligns with resilience-oriented frameworks that link concepts–strategies–examples at neighbourhood level [24] and with residential media architecture studies that operationalize front-yard and interior placements for shared data and collective sense-making [16]. Such placement that is more proximately considered to homes allows media architecture to operate not as an external spectacle but as an integral part of domestic and communal recovery processes, where content purposefully informs, influences, and intrigues rather than distracts [2].

The speculations also reveal that the *users* in the post-disaster contexts should extend beyond traditional audiences of media architecture. Focus group discussions foregrounded residents, particularly in the case for displaced families returning to permanent homes in Antakya, as primary beneficiaries, but also suggested that media architecture must facilitate interaction across multiple stakeholders: architects, local authorities, and even non-human actors such as heritage buildings and stray animals. This broadening of the user definition resonates with more-than-human approaches [28] and with interior media architecture perspectives that treat interiors as socio-technical hubs shaped by social media practices and everyday data [17]. In historic urban fabrics like Antakya, where heritage is a co-agent, projection-based, minimally invasive deployments can stage the *longue durée* of the site while supporting accessibility and inclusion [32].

Another significant point in the focus group sessions underlined the importance of *interaction modalities*. Participants envisioned interfaces that foreground multi-sensory and embodied experiences, such as sound, olfactory, and tactility, rather than relying solely on screens. Prior work criticizes ocular-centrism and argues for full-body, playful engagement [7, 20], while digital placemaking demonstrates how shared and re-usable media infrastructures (LED façades, projection, embedded audio) can scaffold participation and community expression over time [14]. From a practice standpoint, design tools and approaches (prototyping, 3D visualization, evaluation) help match modality-to-context, address robustness, and integrate with physical surroundings that are critical in post-disaster sites.

Taken together, these insights suggest that in post-disaster contexts, media architecture should work through intimate scales, embedded placements, and multisensory interaction, acknowledging more-than-human participants as co-actors. By shifting from monumental icons to situated infrastructures of resilience, media architecture becomes capable of sustaining recovery while remaining attentive to Antakya's social, ecological, and heritage-specific complexities.

**5.2 Understanding implications towards the role of media architecture in post-disaster permanent homes**

The implications of our findings can be understood through the three dimensions we identified: First, facilitating social connection calls for micro-civic interfaces at courtyards, thresholds, and façades that support communication and coordination among residents. This echoes existing evidence that small-scale, shared media displays foster neighbourhood-level sense-making [16]. Second, enabling multispecies participation suggests designing robust, low-tech systems that accommodate ecological actors and material environments, aligning with resilience strategies that 'give non-humans a voice' and acknowledge their agencies [24]. Third, mediating heritage preservation and revival involves non-invasive, sensory media layers that sustain memory, ritual, and identity, as shown in projection-based heritage mediation studies [32]. Altogether, these findings position media architecture in post-disaster permanent homes not as a spectacular addition, but as an embedded socio-technical infrastructure of everyday recovery.

Building on these implications, media architecture also offers a pragmatic and practice-oriented complement to post-disaster recovery frameworks. Prior studies show how the aesthetics of participation enable multisensory engagement and collective sense-making [9], while DIY approaches emphasize how communities can shift from passive audiences to active initiators of change [5]. In Antakya, this means envisioning courtyard projections that residents can appropriate for neighbourhood coordination, or facades that carry layers of memory and ritual without overwhelming their everyday use. In this way, media architecture evolves into a lived and adaptive infrastructure for recovery, continually reshaped by residents as needs, ecologies, and collective identities evolve.

To extend the discourse of post-disaster housing recovery, frameworks such as Build-Back-Better (BBB) and People-Centered Housing Recovery (PCHR) underline that rebuilding process should not replicate pre-disaster vulnerabilities but should foster safer, more holistic environments [21, 22]. PCHR approach in rebuilding process particularly prioritizes resident agency, participation, and culturally grounded practices that also strongly echoed in our Antakya focus groups, where speculative ideas emphasized stewardship, ecological participation, and heritage continuity as integral to permanent homes.

**6 CONCLUSION**

This research examined the role of media architecture in post-disaster permanent housing, drawing on speculative ideas of the focus groups in Antakya after the 2023 earthquakes. Participants envisioned digital and spatial interventions as part of everyday domestic and communal life, intersecting with resilience, participation, and heritage continuity. From these discussions, three dimensions emerged: facilitating social connection through micro-civic interfaces at thresholds, courtyards, and facades; enabling multispecies participation and collective stewardship by extending interaction to a more inclusive landscape; and mediating heritage preservation and revival through non-invasive, participatory, and sensitive media that sustain memory and identity.

These dimensions generate clear design implications that foregrounds disaster as an active agent of design parameters: content must degrade gracefully, placements must remain accessible across shifting phases, and governance must adapt as neighbourhoods repopulate. By acknowledging disaster-agency alongside social and heritage concerns, media architecture is re-positioned as a situated infrastructure of resilience rather than a temporary spectacle, capable of evolving with the rhythms of recovery. Media architecture should prioritize intimate scales and embedded placements, rely on durable and multisensory modalities, and support participatory governance and resident curation. In practice, this could mean adaptable courtyard projections for neighbourhood coordination, appropriable facade displays for storytelling and ritual, and feedback systems responsive to ecological and heritage dynamics. More broadly, these insights show how media architecture can complement frameworks such as Build Back Better and People-Centered Housing Recovery, translating their principles of safety, participation, and cultural grounding into situated, everyday practices of recovery.

To sum up, post-disaster permanent housing should not be seen as standardized shelter provision but as participatory platforms for resilience, acknowledging the agency of the buildings as communicative actors. Media architecture, that is intimate in scale, durable in modality, participatory in governance, and embedded in heritage-sensitive design, can help realize this shift by weaving social, ecological, and cultural recovery into the everyday fabric of rebuilt permanent homes. Future work should investigate long-term appropriation, intergenerational memory, and participatory methods with broader stakeholder engagements, while extending inquiry across other disaster-affected regions to test how media architecture can serve as a situated infrastructure of resilience.